\documentclass[11pt]{article}
\usepackage{amsmath,amssymb,amsfonts,epsfig,graphicx,euscript}%
\newcommand{\be}{\begin{equation}}
\newcommand{\ee}{\end{equation}}

\newcommand{\bse}{\begin{subequations}}
\newcommand{\ese}{\end{subequations}}
\newcommand{\bea}{\begin{eqnarray}}
\newcommand{\eea}{\end{eqnarray}}
\newcommand{\ba}{\begin{array}}
\newcommand{\ea}{\end{array}}

\begin{document}
\hfill%
\vbox{
    \halign{#\hfil        \cr
           IPM/P-2011/012\cr
                     }
      }
\vspace{1cm}
\begin{center}
{ \Large{\textbf{Non-commutative Holographic QCD}\\
\textbf{and}\\
\textbf{Jet Quenching Parameter}\\}} \vspace*{2cm}
\begin{center}
{\bf M. Ali-Akbari}\\%
\vspace*{0.4cm}
{\it {School of Physics, Institute for Research in Fundamental Sciences (IPM)\\
P.O.Box 19395-5531, Tehran, Iran}}  \\
{E-mails: {\tt aliakbari@theory.ipm.ac.ir}}\\%
\vspace*{1.5cm}
\end{center}
\end{center}

\vspace{.5cm}
\bigskip
\begin{center}
\textbf{Abstract}\\
\end{center}
Using gauge/gravity duality, we compute jet quenching parameter in confined and deconfined phases of noncommutative Sakai-Sugimoto model. In the confined phase jet quenching parameter is zero and noncommutativity does not affect it. In deconfined phase we find
that the leading correction is negative \textit{i.e.} it reduces the magnitude of the jet quenching parameter as
compared to its value in commutative background. Moreover it is seen that the effect of leading correction is more pronounced at high temperatures.
\newpage %

%
%
%
%

%
%
%
%
%
%
%

\section{Introduction}
At the Relativistic Heavy Ion Collider (RHIC) at Brookhaven National
Laboratory \cite{RHIC}, quark-gluon plasma (QGP) has been  produced
by Au-Au collision at an ultra-relativistic  center of mass
$\sqrt{s}=200$ Gev. QGP is a low viscosity, hot strongly coupled
fluid. When a very energetic quark or gluon, with transverse momenta
about a few tens Gev is moving through the strongly coupled
plasma it interacts strongly with the plasma and loses energy via
medium-induced radiation so that its transverse momenta becomes at
most about 20 Gev \cite{EXRHIC}. The energy loss of the quark or gluon
in the plasma is described by the jet quenching parameter $\hat{q}$ which
is a property of the strongly coupled QGP. Jet quenching
parameter decreases as the hot plasma expands and cools and the system
re-enters to the confined phase and hadornization takes place. The
time-averaged jet quenching parameter determined in comparison with
RHIC data was found to be around 5-15 Gev$^2$/fm
\cite{Eskola,Dainese}.

Quantum chormodynamic (QCD) is a theory describing strong
interaction between quarks and gluons. The coupling constant of QCD
is large and the interaction is then strong at low energy or large
distance. In other words, at low energy quarks or gluons are not
free and we see colorless combinations of them known as mesons or
hadrons. This is confinement phase where perturbative analysis dose
not work. However at high energy the coupling constant is small and
quarks or gluon can freely move known as deconfinement phase. QGP is
a new phase of QCD. In this new phase the quarks and gluons are
neither confined nor free but instead form some new kind of strongly
interacting fluid. In fact QGP produced at RHIC must be described by
QCD in a regime of strong, and hence nonperturbative, interactions.
As a result, due to strong interactions, QCD perturbative
calculation is not reliable in this phase and it seems that a new
theoretical tools is needed. One of the best candidates is Anti-de
Sitter/Conformal Field Theory (AdS/CFT) correspondence
\cite{Maldacena}.

According to the AdS/CFT correspondence, IIB string theory on
$AdS_5\times S^5$ background is dual to $D=4\ {\cal{N}}=4\ SU(N_c)$
super Yang-Mills theory (SYM) \cite{Maldacena,Aharony}. In the large
$N_c$ and large 't Hooft coupling limit, the SYM theory is dual to
IIb supergravity which is low energy effective theory of superstring
theory. In this setting the strongly coupled SYM at non-zero temperature
corresponds to supergravity in an AdS black brane background where
the SYM theory temperature is identified with the Hawking
temperature of the AdS black hole \cite{Witten}. The AdS/CFT correspondence has
been applied to study different quantities in condense matter
\cite{AliAkbari3} or in  strongly coupled plasma such as shear
viscosity , jet quenching parameter \cite{Liu} and potential between
quark-antiquark system \cite{AliAkbari2}. On the gauge theory side,
the value of jet quenching parameter is related to thermal
expectation value of Wilson loop \cite{Wilson}. On the gravity side,
the related calculation was originally introduced in \cite{Liu} and it
was then elaborated in \cite{D'Eramo}. The result of \cite{D'Eramo}
shows that there is only one extremized string world sheet that is
bounded by light-like Wilson loop and it is the one which stretches from
horizon to boundary. Our aim in this paper is to study jet quenching parameter,
by employing AdS/CFT tools, in the noncommutative
Sakai-Sugimoto model which will be reviewed in next section. Some
properties of this background have been studied in \cite{AliAkbari}.

Non-commutative gauge theories naturally appear on the D-branes with
a background  NSNS $B$-field on them. Explicitly consider a system
of D$p$-branes with a constant NSNS $B$-field along their
worldvolume directions. By taking a certain low energy limit , closed
strings decouple and the resulting action for open strings is the
non-commutative gauge theory \cite{Ardalan}. It is possible to
extend the AdS/CFT dictionary to the cases involving background
$B$-field in the gravity side and noncommutative gauge theory in the
CFT side \cite{Hashimoto:1999ut,Maldacena:1999mh,Alishahiha}. In
addition non-commutativity can be considered as a way to model the magnetic
fields in real system like heavy-ion collisions at RHIC
\cite{Kharzeev} or in QCD \cite{Gusynin}.

In section 2 the noncommutative Sakai-Sugimoto model will be
reviewed. In the next section we will then introduce and compute jet
quenching parameter at low and high temperature noncommutative
backgrounds. The last section is devoted to summary.
In appendix we will calculate jet quenching parameter in
commutative Sakai-Sugimoto model for a more general string configuration.

\section{Noncommutative Sakai-Sugimoto Model}
Holographic QCD background (Sakai-Sugimoto model), at low and high
temperatures, is introduced in  \cite{Aharony2,Sakai2}. This model
is constructed from the near horizon limit of a set of $N_c$
D4-branes compactified on a circle with an anti-periodic boundary
condition for the adjoint fermions. This makes the adjoint fermions
massive and breaks supersymmetry.  In the probe limit, namely $N_f\ll
N_c$ where flavour branes do not backreact on the background,
fermions in (anti-)fundamental representation are introduced by
$N_f$ (anti-)D8-branes intersecting the D4-brane at a 3+1
-dimensional defect. There is thus a global $U(N_f)\times U(N_f)$
chiral symmetry from the worldvolume of D4-brane point of view.

It was shown in \cite{Aharony2} that this theory undergoes a
confinement-deconfinement phase transition at a temperature
$T_d=1/2\pi {\cal{R}}$ where ${\cal{R}}$ is radius of
compactification. For quark separation obeying $R>0.97{\cal{R}}$,
the chiral symmetry is restored at this temperature but for
$R<0.97{\cal{R}}$ there is an intermediate phase which is deconfined
with broken chiral symmetry and the chiral symmetry is restored at
$T_{\chi SB}=0.154R$. In the next two subsections, noncommutative
low and high temperature sakai-Sugimoto model are introduced.

\subsection{Noncommutative low temperature background}
The low temperature noncommutative background is given by
\cite{Alishahiha}
\be\begin{split}\label{Nmetric1} %
 ds^2&=
 (\frac{u}{R})^{3/2}\Big(h(dt_E^2+dx_1^2)+dx_2^2+dx_3^2+fdx_4^2\Big)+(\frac{R}{u})^{3/2}\Big(\frac{du^2}{f}+u^2d\Omega_4^2\Big),\cr
 f&=1-\frac{u_k^3}{u^3},\ \ h=\frac{1}{1+\theta^3u^3},\ \ e^{\phi}=g_s(\frac{u}{R})^{3/4}\sqrt{h},\ \ B\equiv
 B_{t1}=(\frac{\theta}{R})^{3/2}u^3h,\cr
 F_4&=dC_3=\frac{2\pi N_c}{V_4}\epsilon_4,
\end{split}\ee %
where $t_E\ ({\rm{Euclidean\  time}}),\ x^i(i=1,2,3)$ and $x_4$ are
the directions along which the D4-branes are extended. $d\Omega_4^2,
\epsilon_4$ and $V_4=8\pi^2/3$ are the line element, the volume form
and the volume of a unit $S^4$, respectively. $R$ and $u_k$ are
constant parameters. $R$ is related to the string coupling $g_s$ and
string length $l_s$ as $R^3=\pi g_sN_cl_s^3$. $\theta$ is
noncommutative parameter.

The coordinate $u$ is bounded from below by condition $u\geq u_k$.
In order to avoid singularity at $u=u_k$, $x_4$ must be a periodic
variable with period ${\cal{R}}$ \textit{i.e.}%
\be %
 2\pi {\cal{R}}=\frac{4\pi}{3}\sqrt{\frac{R^3}{u_k}}.
\ee %

\subsection{Noncommutative high temperature background}
The high temperature noncommutative background, in units where $R$ is one,
is given by
\be\label{Nmetric2}\begin{split} %
ds^2&=
 u^{3/2}\Big(h\big(fdt_E^2+dx_1^2\big)+dx_2^2+dx_3^2
 +dx_4^2\Big)+u^{-3/2}\Big(\frac{du^2}{f}+u^2d\Omega_4^2\Big),\cr
 f&=1-\frac{u_h^3}{u^3}.
\end{split}\ee %
In this case we have a black hole solution where $u_h$ shows its
horizon. As in low temperature case, at $u=u_h$ we must have %
\be %
u_h=(\frac{4\pi T}{3})^2,
\ee %
where $T$ is background temperature.

\section{Jet Quenching parameter}
The collision of high energy particles can produce jets of
elementary particles. In the initial stage of nucleus-nucleus
collision such as Au-Au at RHIC or Pb-Pb at large hadron collider (LHC)
create QGP and these jets interact with the strongly coupled
medium. Jets then lose energy by radiating gluons as they interact
with the medium (this is similar to bremsstrahlung phenomena in quantum
electrodynamics). This energy loss is called "jet quenching" phenomenon.

\begin{figure}[ht]
\centerline{\includegraphics[width=3.3in]{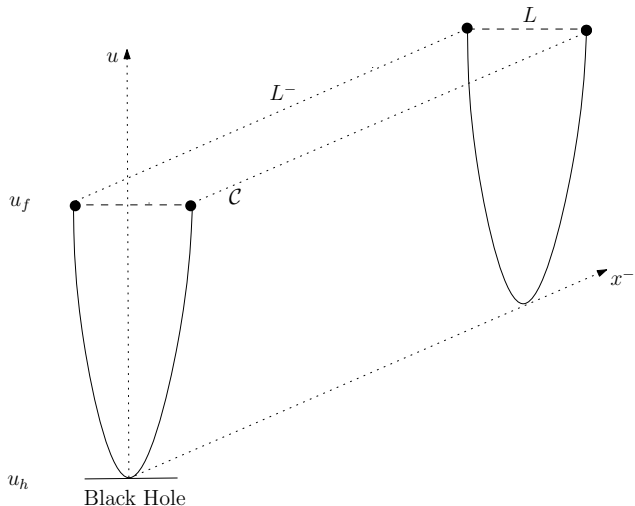}}%
\caption{\label{JQQ} Schematic picture of an open string (describing
quark-antiquark system) moving along $x^-$.}
\end{figure}

As it was argued in  \cite{Kovchegov,Kovner,Kovner2}, the jet quenching can be
quantified through the "jet quenching parameter" $\hat{q}$, which is related to
the thermal expectation value of Wilson loop as follows
\be\label{a} %
 \langle{}W^A({\cal{C}}_{\rm{Light-Like}})\rangle\approx
 e^{-\frac{1}{4\sqrt{2}}~\hat{q}L^-L^2}.
\ee %
The closed rectangular loop
includes two light-like Wilson lines connected by two short
transverse segments of length $L$ (see Fig. \ref{JQQ}). Quark (gluon)
propagates along the light-cone through the medium of length $L^-$
and it is assumed to be large but finite \textit{i.e.} $L^-\gg L$.
From AdS/CFT corresponding, in the large 't Hooft coupling and large
$N_c$ limits the thermal expectation value of $\langle
W^F({\cal{C}})\rangle$ can be related to the extremal string action
in AdS black hole background. In fact what we have is
\cite{Maldacena,Aharony,Witten} \footnote{In \eqref{a} and
\eqref{f}, $A$ and $F$ denote adjoint and fundamental
representation. For $SU(N_c)$, the Wilson line in adjoint
representation can be obtained using the identity $\rm{Tr} W=\rm{tr}
W \rm{tr} W^\dagger-1$ where $\rm{Tr}$ and $\rm{tr}$ denote trace in
the adjoint and fundamental representation respectively. }
\be\label{f} %
 \langle W^F({\cal{C}})\rangle=e^{-S({\cal{C}})}.
\ee %
It is clearly seen that, in order to compute jet quenching
parameter, we need to find the value of the string action in our
specific background. We will do this computation in the next two
following subsections.

\subsection{Jet quenching parameter in noncommutative low temperature}
The noncommutative background \eqref{Nmetric1} in the light-cone coordinate is given by %
\be\label{LNmetric1}\begin{split} %
ds^2=&
 u^{3/2}\Big(-hdx^+dx^-+dx_2^2+dx_3^2+fdx_4^2\Big)+u^{-3/2}\Big(\frac{du^2}{f}+u^2d\Omega_4^2\Big),
\end{split}\ee %
where $x^\pm=ix_1\pm t_E$. In the new coordinate,
the $B$-field in the background becomes %
\be\label{Bfield} %
 B=B_{+-}=\theta^{3/2}u^3h.
\ee %
The action of string in an arbitrary background with a nontrivial $B$-field is
\footnote{Note that by setting $R=1$ we have $\alpha'=(\frac{1}{\pi g_sN_c})^{2/3}$.} %
\be\label{Laction} %
 S=\frac{1}{2\pi\alpha'}\int d\tau d\sigma\big(\sqrt{-\det
 g_{\mu\nu}}+\epsilon^{\mu\nu}B_{MN}\partial_\mu x^M \partial_\nu x^N\big), %
\ee %
where $g_{\mu\nu}=G_{MN}\partial_\mu x^M \partial_\nu x^N$. $G_{MN}$
is background metric and $M$ index runs from 0 to 9.
$\mu(=\tau,\sigma)$ labels string worldsheet where in static gauge
we set $\tau=x^-$ and $\sigma=x_2$. By choosing $u(\sigma)$ and
constant values for other coordinates, in static gauge, we have %
\be\begin{split} %
 g_{\tau\tau}&=0,\cr
 g_{\tau\sigma}&=0,
\end{split}\ee %
and hence the first term in \eqref{Laction} clearly becomes zero. The
second term in \eqref{Laction} does not contribute in the action
because $x^+$ has been supposed to be a constant. As a result at low
temperature jet quenching parameter vanishes.

By setting $\theta$ equal to zero, \eqref{LNmetric1} reduces to
commutative background describing confining phase of QCD \cite{Aharony2}.
At commutative low energy jet quenching parameter was computed in
\cite{Gao} and reported that it is zero. Then our result means that
noncommutativity does not change the value of jet quenching
parameter at low temperature and it is still zero. Since jet quenching
parameter describes interaction between test quark (gluon) and
strongly coupled medium, vanishing jet quenching parameter tells us
that the medium is transparent to test quark (gluon). This, as expected,
is a sign of being in a confined phase.

\subsection{Jet quenching parameter in noncommutative high temperature}
The noncommutative background \eqref{Nmetric2} in the light-cone coordinate is given by %
\be\label{LNmetric2}\begin{split} %
ds^2=&
 u^{3/2}\Big(-\frac{(1+f)h}{2}dx^+dx^-+dx_2^2+dx_3^2+dx_4^2\Big)\cr
 -&\frac{(1-f)h}{4} u^{3/2}\big[(dx^+)^2+(dx^-)^2\big]
 +u^{-3/2}\Big(\frac{du^2}{f}+u^2d\Omega_4^2\Big).
\end{split}\ee %
By choosing $u(\sigma)$ and constant values for other coordinates
\footnote{Another configuration has been considered in the appendix.}
the final action, in static gauge, is found by substituting
\eqref{LNmetric2} and \eqref{Bfield} in \eqref{Laction}
and then %
\be\label{lagranigian} %
 S=\frac{u_h^{3/2}}{4\pi\alpha'} \int_0^{L^-}d\tau \int d\sigma\sqrt{h(1+\frac{u'^2}{fu^3})},%
\ee %
where $u'=\frac{du}{d\sigma}$. Note that the second term in
\eqref{Laction} does not contribute in the action as in the
pervious case. Light-cone time does not explicitly appear in the
action and we will therefore have a conserved charge,
energy of the system $E$. We have then %
\be %
 E={\cal{L}}-u'\frac{\partial{\cal{L}}}{\partial u'},
\ee %
and by using \eqref{lagranigian}, the above equation leads to  %
\be\label{derivative} %
 E^2u'^2=u^3(h-E^2)f.
\ee %

The above equation can be solved to find $L$ by using appropriate
boundary condition. The boundary condition introduced in \cite{Liu}
is to consider curve ${\cal{C}}$ as a boundary of the string
worldsheet by imposing $u'(\pm \frac{L}{2})=\infty$ which preserve
the symmetry $u(\sigma)=u(-\sigma)$. Moreover, as it was discussed
in \cite{D'Eramo}, the turning point of the string where $u'=0$ is
located at $\sigma=0$ (where $u(\sigma=0)=u_h$) as implied by the symmetry.
In other words, string stretches from infinity to horizon
and then returns to infinity.

As it was discussed in \cite{Maldacena:1999mh} in the AdS/CFT
correspondence, IR limit of the thermal noncommutative gauge theory
can be described by geometry \eqref{Nmetric2}.  On the gravity side
it means that $u$ can not go to infinity and the value of
$\theta^3u^3$ must always be small for a fixed value of
$\theta$. Therefore we set $u(\pm \frac{L}{2})=u_f$ where $u_f$ is the
upper limit of the $u$ coordinate.
Hence \eqref{derivative} leads to %
\be\label{tol} %
 \frac{L}{2}=\int_0^{L/2}d\sigma=E\int_{u_h}^{u_f}\frac{du}{\sqrt{(h-E^2)(u^3-u_h^3)}}.
\ee %
By expanding the right hand side of \eqref{tol} up to $O(\theta^6)$
terms, we have \footnote{Our expansion is reliable when $\theta\int d\sigma u\ll1$.}
\be\begin{split}\label{tol2} %
 \frac{L}{2E}&=\frac{1}{\sqrt{1-E^2}}\Big(\int_{u_h}^{u_f}\frac{du}{\sqrt{u^3-u_h^3}}
 -\frac{\theta^3}{2(1-E^2)}\int_{u_h}^{u_f}\frac{u^3du}{\sqrt{u^3-u_h^3}}\Big).
\end{split}\ee %
As it was mentioned earlier, since gravity theory with B-field
describes IR region of the gauge theory \cite{Maldacena:1999mh}, $u$
can not go to infinity and for this reason we can consider
$u_f=bu_h$ and hence \eqref{tol2} becomes%
\be\begin{split} %
\frac{L}{2E}&=\frac{1}{\sqrt{1-E^2}}\Bigg(\frac{2\Gamma(\frac{7}{6})\sqrt{b\pi}-\Gamma(\frac{2}{3})
F(\frac{1}{6},\frac{1}{2},\frac{7}{6},\frac{1}{b^3})}{
\Gamma(\frac{2}{3})\sqrt{bu_h}} \cr
&-\frac{\theta^3}{2(1-E^2)}\frac{1}{15}u_h^{5/2}\bigg[\frac{5\Gamma(-\frac{5}{6})\sqrt{\pi}}{\Gamma(-\frac{1}{3})}+6
b^{5/2}F(-\frac{5}{6},\frac{1}{2},\frac{1}{6},\frac{1}{b^3})\bigg]\Bigg),
\end{split}\ee %
where $F$ and $b$ are Gauss hypergeometric function and a finite
constant number. Setting $b=50$, for simplicity, the above equation
is simplified as
\be\label{a-L}%
 L=\frac{4.3a}{\sqrt{u_h}}\Big(1+(1+a^2)\hat{\theta}^3 u_h^3\Big),
\ee %
where $a^2=\frac{E^2}{1-E^2}$ and $\hat{\theta}=1644.63\theta$. Note
that the different value of $b$ changes coefficients of the above
equation. \eqref{a-L} is a third order equation in $a$ and has
therefore
three roots where its real one is given by  %
\be\label{constant} %
 a^2=\frac{L^2u_h}{(4.3)^2}\Big(1-2\big[1+\frac{L^2u_h}{(4.3)^2}\big]\hat{\theta}^3u_h^3\Big).
\ee %
This gives us a relation between $a$ and $L$ which can be used to
simplify the string action.

The string action at high temperature background is given by %
\be\begin{split} %
 S&=\frac{u_h^{3/2}L^-}{4\pi\alpha'E}\int d\sigma ~h \cr
 &=\frac{u_h^{3/2}L^-}{2\pi\alpha'}\int_{u_h}^{u_f}\frac{hdu}{\sqrt{(h-E^2)(u^3-u_h^3)}}.
\end{split}\ee %
Similar to what is done in \eqref{tol2}, one easily finds
\be\begin{split} %
 S&=\frac{u_h^{3/2}L^-}{2\pi\alpha'\sqrt{1-E^2}}\Big(\int_{u_h}^{u_f}\frac{du}{\sqrt{u^3-u_h^3}}
 -\frac{\theta^3(1-2E^2)}{2(1-E^2)}\int_{u_h}^{u_f}\frac{u^3du}{\sqrt{u^3-u_h^3}}\Big)\cr
 &+O(\theta^6),
\end{split}\ee %
and then %
\be\label{nonaction}\begin{split} %
 S&=\frac{2.15u_hL^-}{2\pi\alpha'\sqrt{(1-E^2)}}\Big(1-\frac{1-2E^2}{1-E^2}~\hat{\theta}^3u_h^3\Big).
\end{split}\ee %
Regarding to relation between $a$ and $E$, \textit{i.e.}
$a^2=\frac{E^2}{1-E^2}$, $E$ can be found from \eqref{constant} and
by substituting in \eqref{nonaction}, we have %
\be\begin{split} %
 S&=\frac{2.15L^-u_h}{2\pi\alpha'}\big(1-\hat{\theta}^3u_h^3\big)\sqrt{1+0.05L^2u_h}\cr
 &\simeq\frac{2.15L^-u_h}{2\pi\alpha'}\big(1-\hat{\theta}^3u_h^3\big)(1+\frac{0.05}{2}L^2u_h),
\end{split}\ee %
where in the last line square root was expanded for small transverse distance, $LT\ll1$. %

As was discussed in \cite{Liu}, the above calculation includes
the self-energy of quark-antiquark system. In order to subtract the
self-energy, we consider a trivial configuration given by two
disconnected open strings, each of them descend from $u_f$ to $u_i$
at constant $x_2$, one at $+\frac{L}{2}$ and the other at
$-\frac{L}{2}$. The string action then becomes %
\be\label{Selfenergy}\begin{split} %
 S_s&=\frac{L^-}{\pi\alpha'}\int_{u_h}^{u_f}du\sqrt{g_{--}g_{uu}}\cr
 &=\frac{L^-}{2\pi\alpha'}\int_{u_h}^{u_f}du\sqrt{\frac{h(1-f)}{f}}.
\end{split}\ee %
\begin{figure}[ht]
\centerline{\includegraphics[width=3.3in]{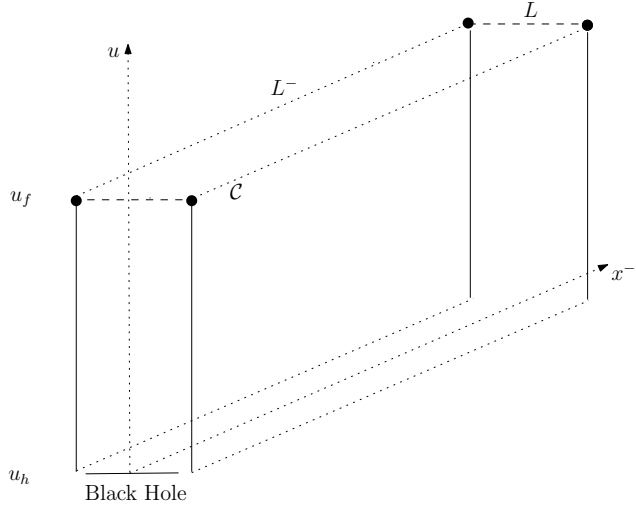}}%
\caption{\label{JQ1} Two disconnected open strings moving along $x^-$. }
\end{figure}%
By substituting form \eqref{Nmetric2} in \eqref{Selfenergy}, we finally have
\be %
 S_s=\frac{2.15u_hL^-}{2\pi\alpha'}\big(1-\hat{\theta}^3u_h^3\big)+O(\theta^6).
\ee %
The total action is given by %
\be %
 S_t\equiv
 S-S_s=\frac{0.03}{\pi\alpha'}\big(1-\hat{\theta}^3u_h^3\big)L^2L^-u_h^2.
\ee %
Footnote 1 implies that in the large $N_c$ limit
\be %
 \langle{}W^A({\cal{C}}_{\rm{Light-Like}})\rangle=2\langle{}W^F({\cal{C}}_{\rm{Light-Like}})\rangle,
\ee %
and the adjoint Wilson loop \eqref{a} is then given by ${\rm{exp}}(-2S_t)$. Jet quenching parameter finally is %
\be\begin{split} %
 \hat{q}&=\frac{0.06}{\pi\alpha'}\big(1-\hat{\theta}^3u_h^3\big)u_h^2\cr
 &=7.35\bigg(1-(\frac{4\pi}{3})^6\hat{\theta}^3T^6\bigg)\lambda T^4.
\end{split}\ee %
where $\lambda=4\pi g_sN_c\alpha^{'1/2}$. Setting $\theta=0$, \textit{i.e.} commutative case, jet quenching parameter is proportional to $\lambda$ and $T^4$, as reported in \cite{Gao}. Moreover in the noncommutative background we see a decrease in value of jet quenching parameter. However this reduction seems to be more important at high temperature, we should notice that the above result has been obtained assuming  $\hat{\theta}u_h \ll 1$. Because this term is the first term of an expansion in terms of $\theta$.

\section{Summary }

Jet quenching is one of the most important characteristic features of nuclear matter in the quark-gluon plasma phase, which has been observed at RHIC or heavy ion collisions at LHC. In this paper the effect of noncommutativity
on the jet quenching parameter was studied at low and high temperatures of noncommutative Sakai-Sugimoto background.

At low temperature case where the system is in the confined phase, we expect the jet quenching phenomenon to be absent as the medium is transparent to the gluon or quark jets. In other words there is no interaction between test quark or gluon and the medium.

At high temperature case the effect of noncommutativity reduces the magnitude of the jet quenching parameter as compared to its value in the commutative case.

\section{Acknowledgment}
We are grateful to thank M. M. Sheikh-Jabbari for useful discussions
and comments. We also like to thank K. Bitaghsir for helpful
discussion.
\appendix
\section{A more general string configuration and jet quenching parameter }
Here we consider a more general configuration but in commutative
high temperature background. In static gauge by choosing $u(\sigma)$,
$x_4(\sigma)$ and constant values for other coordinates the string
action becomes %
\be\label{lagranigian2} %
 S=\frac{u_h^{3/2}}{4\pi\alpha'} \int_0^{L^-}d\tau \int d\sigma\sqrt{1+x^{\prime 2}_4+\frac{u'^2}{fu^3}},%
\ee %
where $u'=\frac{du}{d\sigma}$ and $x'_4=\frac{dx_4}{d\sigma}$. Note
that in commutative case the second term in \eqref{Laction} is zero. As before light-cone time does not
explicitly appear in the action and energy of the system $\hat{E}$ is a constant which is given by %
\be\label{a1} %
 \hat{E}={\cal{L}}-u'\frac{\partial{\cal{L}}}{\partial u'}-x'_4\frac{\partial{\cal{L}}}{\partial x'_4},
\ee %
and then %
\be %
 \hat{E}=\frac{h}{\sqrt{h(1+x^{\prime 2}_4+\frac{u'^2}{fu^3})}}.
\ee %
Equation of motion for $x_4$ is  %
\be %
 x'_4=\frac{C}{\hat{E}},
\ee %
where $C$ is a constant and as a result $x'_4$ is also a constant.
From \eqref{a1} it is easy to find a relation among $\hat{E}$, $C$
and $L$ defined in \eqref{tol} which is %
\be %
 L=\frac{4.84\hat{a}}{\sqrt{u_h}},
\ee %
where $\hat{a}^2=\frac{E^2}{1-tE^2}$ and
$t=1+(\frac{C}{\hat{E}})^2$.
The string action then becomes %
\begin{figure}[ht]
\centerline{\includegraphics[width=2in]{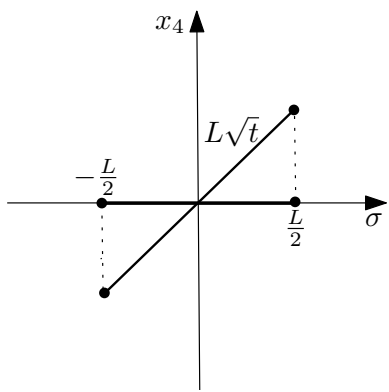}}%
\caption{\label{JQ2} The difference between physical distance $L\sqrt{t}$ and $L$ has been shown. }
\end{figure}%
\be\begin{split} %
 S\simeq\frac{2.42L^-u_h}{2\pi\alpha'}(1+0.02L^2tu_h),
\end{split}\ee %
$L^2t$ is a new parameter appearing in the jet quenching parameter.
As it is shown in Fig. \ref{JQ2}, this quantity is physical distance which
must appear in physical quantity. In fact the equation of motion for
$x_4$ states that $\frac{C}{\hat{E}}$ can be zero or constant. If $x_4=0$, $L$ is physical distance and otherwise our results are written
in terms of physical distance $L^2t$.

Obviously, after subtracting the self-energy of the system, jet
quenching parameter becomes %
\be\begin{split} %
 \hat{q}=\frac{0.04}{\pi\alpha'}u_h^2.
\end{split}\ee %

\end{document}